\documentclass[a4paper]{article}
\usepackage{INTERSPEECH2022}
\usepackage[dvipsnames]{xcolor}
\usepackage{amsmath,graphicx}
\usepackage{adjustbox}
\usepackage{multirow}
\usepackage{float}
\usepackage{booktabs}
\usepackage{array}
\usepackage{url}
\usepackage{algorithm}
\usepackage{algpseudocode}
\usepackage{enumitem}
\usepackage{cite}
\usepackage{hyperref}
\usepackage{cleveref}
\usepackage{multirow}
\usepackage{comment}
\usepackage{makecell}

\makeatletter
\newcommand*\bigcdot{\mathpalette\bigcdot@{.5}}
\newcommand*\bigcdot@[2]{\mathbin{\vcenter{\hbox{\scalebox{#2}{$\m@th#1\bullet$}}}}}
\makeatother

\definecolor{myorangedark}{RGB}{201, 92, 46}
\definecolor{myreddark}{RGB}{151,79,75}
\definecolor{mybluedark}{RGB}{48,112,183}
\definecolor{mylightgreen}{RGB}{88,189,182}

\title{Unsupervised Voice Activity Detection by Modeling Source and System Information using Zero Frequency Filtering}
\name{Eklavya Sarkar$^{1,2}$, RaviShankar Prasad$^{1}$, Mathew Magimai.-Doss$^1$
}
\address{
  $^1$Idiap Research Institute, Martigny, Switzerland\\
  $^2$Ecole polytechnique fédérale de Lausanne, Switzerland}
\email{\{eklavya.sarkar, ravi.prasad, mathew\}@idiap.ch}

\begin{document}

\maketitle
\begin{abstract}
Voice activity detection (VAD) is an important pre-processing step for speech technology applications. The task consists of deriving segment boundaries of audio signals which contain voicing information. In recent years, it has been shown that voice source and vocal tract system information can be extracted using zero-frequency filtering (ZFF) without making any explicit model assumptions about the speech signal. This paper investigates the potential of zero-frequency filtering for jointly modeling voice source and vocal tract system information, and proposes two approaches for VAD. The first approach demarcates voiced regions using a composite signal composed of different zero-frequency filtered signals. The second approach feeds the composite signal as input to the rVAD algorithm. These approaches are compared with other supervised and unsupervised VAD methods in the literature, and are evaluated on the Aurora-2 database, across a range of SNRs ($20$ to $-5$ dB). Our studies show that the proposed ZFF-based methods perform comparable to state-of-art VAD methods and are more invariant to added degradation and different channel characteristics. 
\end{abstract}
\noindent\textbf{Index Terms}: Voice activity detection, zero-frequency filtering, speech analysis, signal processing.

\section{Introduction}
Voice activity detection (VAD) refers to the task of identifying segment boundaries in audio signals which essentially contain voicing information, and typically is one of the first steps to be carried out in any speech technology. Computational efficiency and robustness to noisy data are thus essential pre-requisites for any state-of-the-art voice activity detector. VAD methods can be broadly categorised as unsupervised and supervised methods.

Early unsupervised methods made use of simple energy-based features and temporal parameters such as zero-crossing rate (ZCR)~\cite{Rabiner1975AnAF, 1163642}, before applying a discriminator model to compute the speech/non-speech decision boundary. Spectral features based on autocorrelation \cite{autocorr_robust, Kristjansson2005VoicingFF, 6403507}, mel-frequency cepstral coefficients (MFCCs) \cite{Kristjansson2005VoicingFF}, skewness and kurtosis of linear prediction (LP) residual~\cite{905996}, spectral shape \cite{1162964}, harmonic structure~\cite{vad_periodicity_measure}, voicing \cite{5704566}, cepstral features~\cite{Haigh93avoice, 327987}, perceptual spectral flux \cite{sadjadi2013unsupervised}, spectral flatness (SF) and short-term energy \cite{7077834}, and speech enhancement and denoising through pitch indicators \cite{TAN20201} were proposed to improve the performance and robustness of these systems in the presence of noise. 

Supervised VAD models mostly rely on a likelihood ratio test (LRT) over the estimated parameters in a maximum likelihood (ML) framework \cite{sohn1998voice}. Recent deep learning models have also shown success. Deep belief networks \cite{zhang2012deep} combine various acoustic features through multiple non-linear hidden layers to discover the manifold of the features and observe regularity among them to predict the frame class. Hybrid deep architectures, incorporating convolutional neural networks (CNNs) and long short-term memory (LSTM) models, based on raw waveform \cite{ZazoIS16} and spectrograms \cite{cnnbilstm_vad}, have also been proposed. Although these methods can yield good performance, they come at a high computational cost, requiring training or fine-tuning a pre-trained model, ground truth labels, and do not rely as much on prior knowledge as unsupervised methods.

The focus of this paper lies on unsupervised VAD methods, which tend to incorporate prior knowledge about voice source and vocal tract system, typically through source-system decomposition methods, such as linear prediction analysis and cepstral analysis. Unlike such methods, which make a mathematical model assumption, it has been shown in recent years that voice source and vocal tract system information can be effectively extracted using zero-frequency filtering \cite{4648930, prasad21_interspeech} without making any such explicit model assumptions about the speech signal. This paper investigates the potential of zero-frequency filtering for jointly modeling voice source and vocal tract system system information to perform VAD. To that end, we demonstrate that voice activity detection can be effectively achieved by combining the outputs of a bank of zero-frequency filters that carry information related to the fundamental frequency ($f_0$), and the first ($F_1$) and second ($F_2$) formants. 

The remainder of the paper is organized as follows. \Cref{sec:background} provides an overview on the background of zero-frequency filtered signals. \Cref{sec:method} presents the two approaches for VAD using the ZFF signals. \Cref{sec:experiments} gives the experimental setup used to validate our method, and \Cref{sec:results} summarizes the results. Finally, \Cref{sec:conclusion} concludes the paper.

\begin{figure*}[!htb]
  \centering
  \includegraphics[width=\linewidth]{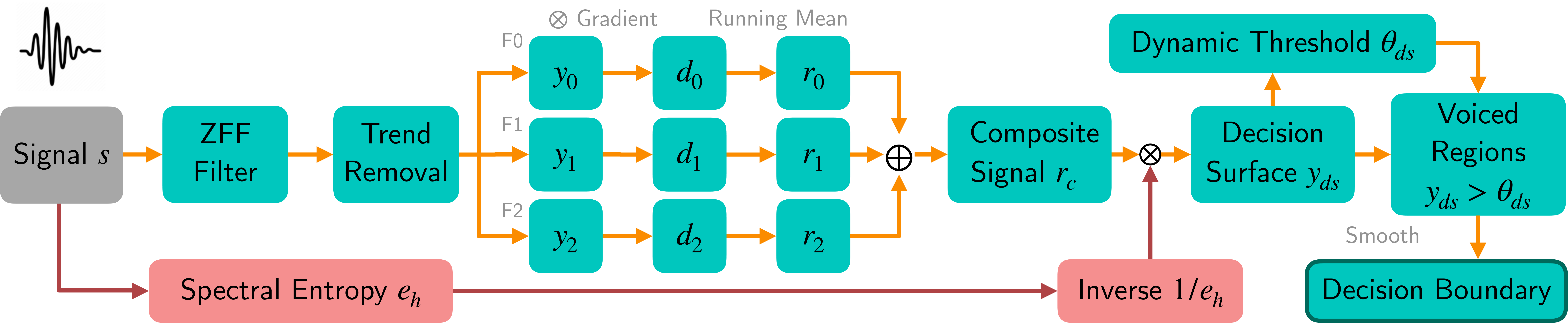}
  \caption{Complete pipeline of proposed method to derive a decision boundary for voice activity detection.}
  \label{fig:zff_pipeline}
\end{figure*}

\section{Background} \label{sec:background}

Zero-frequency filtering (ZFF) was originally proposed in the context of extracting information related to voice source~\cite{4648930}. In this method, a speech signal $s$ is first passed through cascaded digital resonators, implemented as an integrator, centered at $0$ Hz, i.e. a zero-frequency filter. The resulting impulse response is given in \cref{eq:cascade} and its equivalent transfer function in \cref{eq:transfer_func}.
\begin{equation}
    x[n] = s[n] - 2x[n-1] + x[n-2] \label{eq:cascade}
\end{equation}
\begin{equation}
     H[z] = \frac{1}{1-2z^{-1}+z^{-2}} \label{eq:transfer_func}
\end{equation}
A trend removal (i.e. local mean subtraction) step, based on an estimate of the periodicity of the speech signal, is then applied to the output of the cascaded resonators to obtain glottal closure instance (GCI) locations and strength of excitation information. The trend removal operation is described by \cref{eq:trend_rem}.
\begin{equation}
    y[n] = x[n] - \frac{1}{2N+1} \sum^{n+N}_{k=n-N}x[k]; N+1 \leq n \leq L-N. \label{eq:trend_rem}
\end{equation}
$L$ corresponds to the length of the signal and $2N+1 \sim T_0$ is the trend removal window duration. $T_0$ is estimated through autocorrelation.

In a recent work, it was shown that, although the zero-frequency filter heavily damps the high frequency regions in the speech signal, $F_1$ and $F_2$ information can still be estimated by moderating the trend removal window duration~\cite{prasad21_interspeech}.

\begin{figure}[!htb]
\begin{minipage}{\linewidth}
  \centering
  \includegraphics[width=\textwidth]{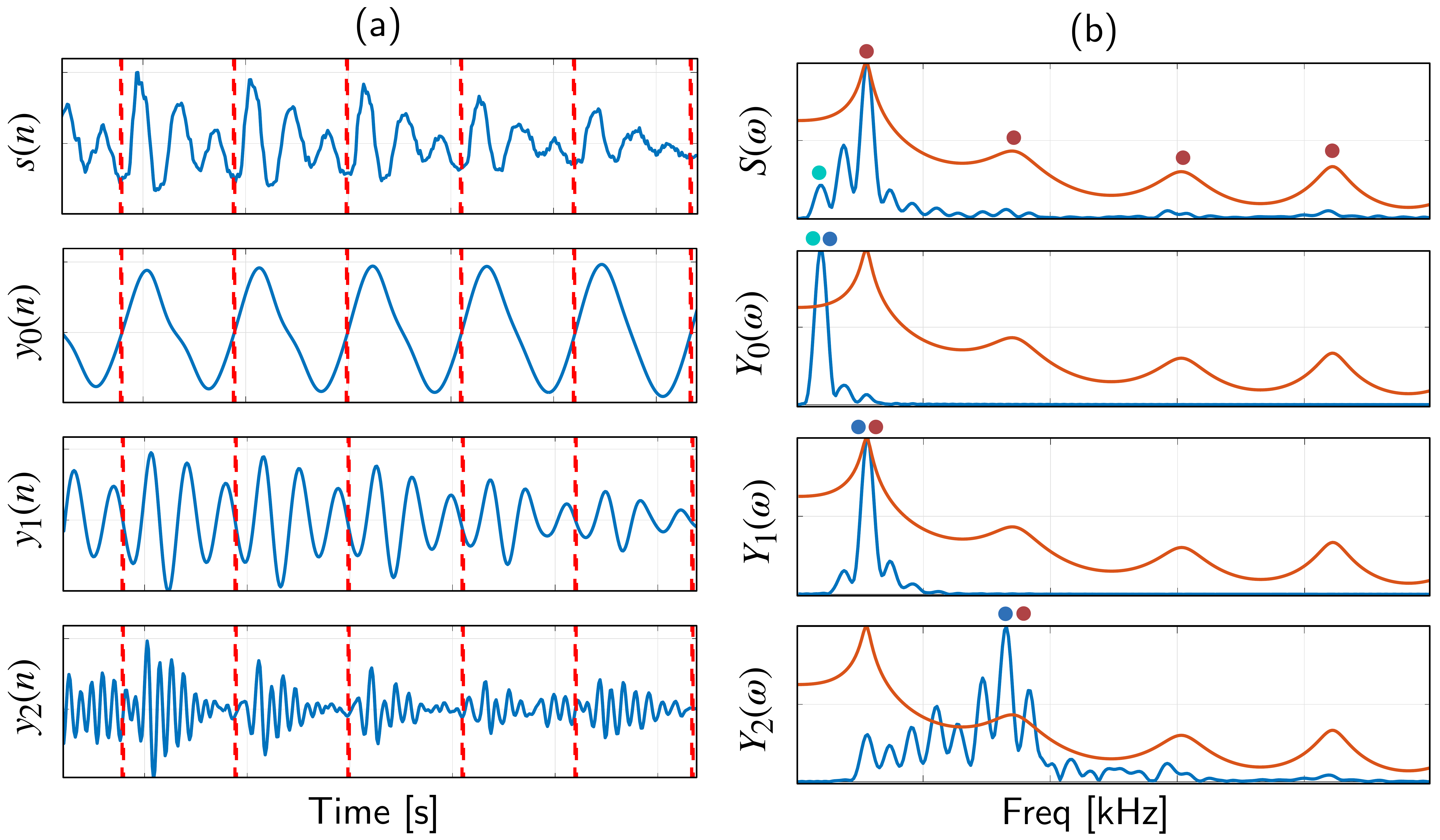}
  \caption{(a1) Speech signal. (a2--a4) ZFF signals $y_0(n)$, $y_1(n)$, $y_2(n)$. GCI locations (\textcolor{red}{\textbf{- -}}). (b1) $S(\omega)$ (\textcolor{mybluedark}{\textbf{---}}) and its envelope (\textcolor{myorangedark}{\textbf{---}}). Formant peaks (\textcolor{myreddark}{$\bullet$}). Fundamental frequency peak (\textcolor{mylightgreen}{$\bullet$}). (b2--b4) $Y_0(\omega)$, $Y_1(\omega)$, $Y_2(\omega)$ (\textcolor{mybluedark}{\textbf{---}}), and respective peaks (\textcolor{mybluedark}{$\bullet$}).}
  \label{fig:zff_dft_plot}
\end{minipage}
\end{figure}

\Cref{fig:zff_dft_plot} illustrates the extraction of $f_0$, $F_1$, and $F_2$ evidences. A voiced speech signal $s(n)$ and the location of glottal closure instants (GCIs) (\textcolor{red}{\textbf{- -}}) are presented in (a1). The latter is derived from the negative-to-positive zero-crossing locations using the method in \cite{4648930}. The corresponding discrete Fourier transform (DFT) spectrum $S(\omega)$ (\textcolor{mybluedark}{\textbf{---}}) and the inverse filter response (\textcolor{myorangedark}{\textbf{---}}) obtained through LP analysis are given in (b1). The fundamental frequency value is obtained at the first peak (\textcolor{mylightgreen}{$\bullet$}) of the DFT spectrum $S(\omega)$. The formants can also be located at global and relative peak locations  (\textcolor{myorangedark}{$\bullet$}) in the spectral envelope. The signals $y_0(n)$, $y_1(n)$, $y_2(n)$ and the corresponding GCI locations, shown in (a2--a4), are obtained by passing the speech signal through the ZFF filter, and computing a trend removal step with three different trend removal windows, in this case $T_0$, $T_0/5$, $T_0/10$, where the estimate of fundamental period $T_0$ is calculated through autocorrelation. Their respective DFT response $Y_0(\omega)$, $Y_1(\omega)$, $Y_2(\omega)$, and corresponding peaks (\textcolor{mybluedark}{$\bullet$}), as well an overlay of the envelope of $S(\omega)$, are given in (b2--b4).

\section{Proposed Method} \label{sec:method}

As motivated in the previous section, zero-frequency filtered signals effectively encode salient source and system speech information, such as the fundamental frequency $f_0$, and formants $F_1$ and $F_2$. Furthermore, these evidences are also robust to noise, as the SNR is high in the regions where the source and system related information manifests in the time-frequency domain. More precisely, GCIs in the time domain are high SNR regions, and similarly, $F_1$ and $F_2$ appear as peaks in the short-time spectrum. Furthermore, while selectively focusing on the source and system evidences, ZFF heavily damps rest of the spectral information, potentially suppressing other interferences. Considering these aspects, we propose the following two approaches for VAD based on zero-frequency filtering:
\begin{enumerate}
    \item In the first approach, illustrated in \Cref{fig:zff_pipeline}, the outputs of the different zero-frequency filters are combined to obtain a composite signal carrying $f_0$, $F_1$, and $F_2$ related information. Voiced regions are demarcated by applying a spectral-based weighing.
    \item In the second approach, the composite signal is given as the input to another VAD algorithm.
\end{enumerate}
In the remainder of this section, we present the details of the first approach.

\begin{algorithm}[htb]
\caption{Proposed VAD method using ZFF.}
\label{alg:zff_vad}
\begin{algorithmic}
    \State 1. Compute $x[n]$: $x[n] = s[n] \circledast H_Z[n]$, using \cref{eq:transfer_func}.
    \State 2. Estimate $T_0$ (i.e., $1/f_0$) for $s[n]$ using autocorrelation.
    \State 3. Obtain $y_0[n]$, $y_1[n]$, $y_2[n]$ from $x[n]$ using \cref{eq:trend_rem}, with windows sizes: $2N+1 \sim [T_0, T_0/5, T_0/10]$.
    \State 4. Determine $d_0[n]$, $d_1[n]$, $d_2[n]$ by weighing $y_0[n]$, $y_2[n]$, $y_3[n]$ with their gradients: $d_i = y_i[n] \cdot ( y_i[n]-y_i[n-1] )$ to highlight the regions of interest~\cite{kadiri2020excitation, prasad_icassp20}.
    \State 5. Obtain the running mean signal $r_0[n]$, $r_1[n]$, $r_2[n]$ from $d_0[n]$, $d_1[n]$, $d_2[n]$ over a duration of $40$ ms.
    \State 6. Calculate the accumulated signal $r_c[n] = r_0[n] + r_1[n] + r_2[n]$, and normalize it between [$0$--$1$].
    \State 7. Obtain the spectral entropy $e_h[n]$ from $x[n]$ by computing the spectrum with FFT over a window of $20$ ms.
    \State 8. Obtain the decision surface $y_{ds}[n] = r_c[n] \cdot 1/e_h[n]$.
    \State 9. Derive a dynamic threshold every $300$ ms: $\theta_{\text{ds}} = ds_{\min}+ (ds_{\text{med}}/3)$, where $ds_{\min} = \min\{y_{\text{ds}}[n]\}$ and $ds_{\text{med}} = \text{median}\{y_{\text{ds}}[n]$\}.
    \State 10. Demarcate voiced regions as $y_{ds}[n] \geq \theta_{\text{ds}}$.
    \State 11. Smooth decision boundary by eliminating short duration outlier segments and merging those in close proximity.
\end{algorithmic}
\end{algorithm}

\begin{figure}[!htb]
\begin{minipage}{\linewidth}
  \centering
  \includegraphics[width=\textwidth]{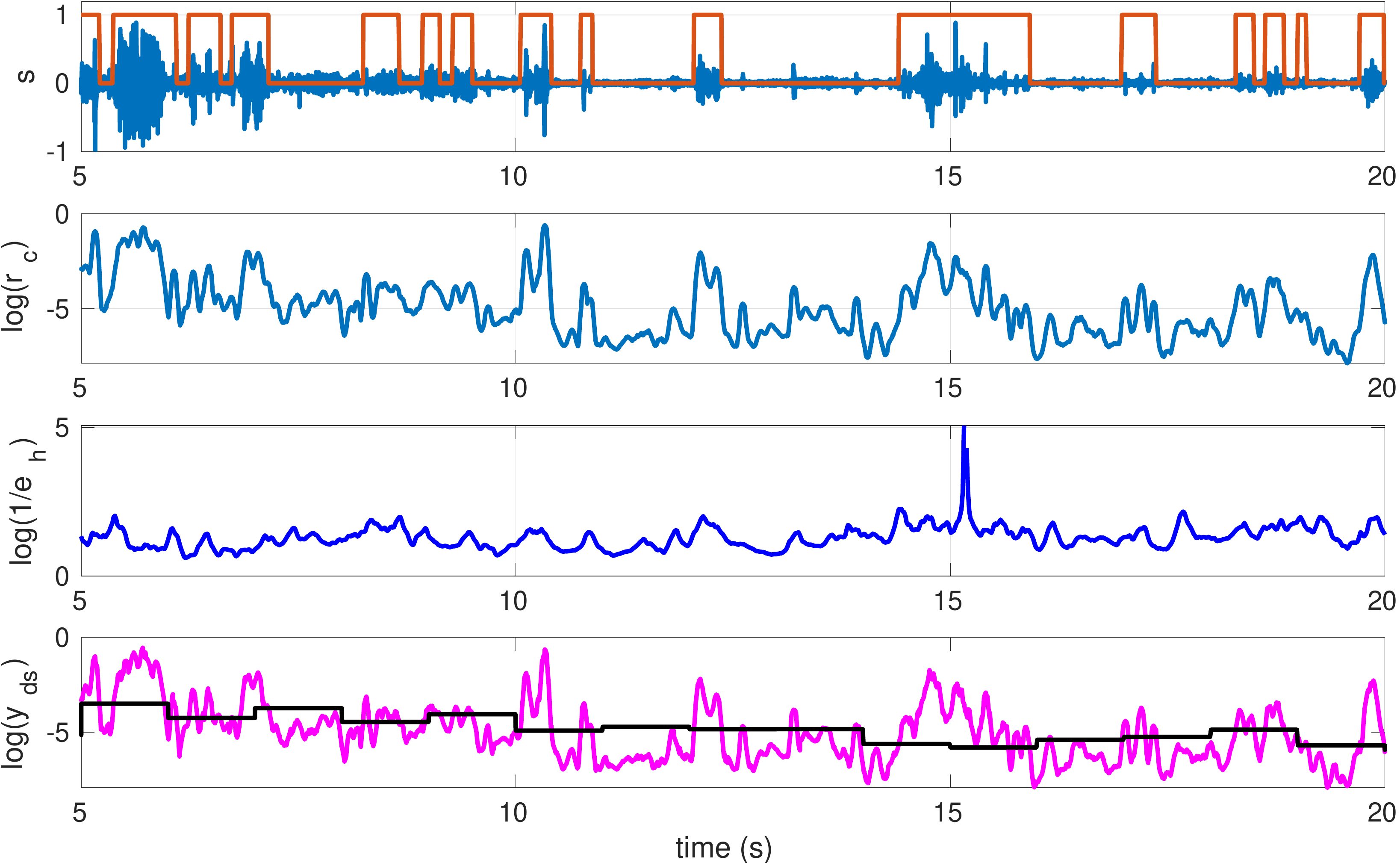}
  \caption{a) Naturally corrupted speech signal $s$ and final decision boundary. b) Accumulated ZFF signals $r_c$ c) Inverse spectral entropy $1/e_{h}$ d) Decision surface $y_{ds}$ and dynamic threshold $\theta_{\text{ds}}$.}
  \label{fig:zff_vad_subplots}
\end{minipage}
\end{figure}

\Cref{alg:zff_vad} presents the proposed VAD method, with successive steps represented in \Cref{fig:zff_pipeline}. \Cref{fig:zff_vad_subplots} illustrates the principal components of this technique. \Cref{fig:zff_vad_subplots} (a) shows a naturally corrupted speech signal along with the boundary demarcations for voiced segments, obtained using the proposed method. \Cref{fig:zff_vad_subplots} (b) shows the composite signal $r_c$ obtained after applying the zero-frequency filtering stage, and \Cref{fig:zff_vad_subplots} (c) the inverse spectral entropy weight that is applied to it. Finally, \Cref{fig:zff_vad_subplots} (d) shows the resulting decision surface based on dynamic threshold estimation every $300$ ms. The duration is chosen on the heuristic assumption that the baseline for the decision surface will not significantly change within this period.


\section{Experimental Setup} \label{sec:experiments}

This section presents the dataset, baseline methods, and the evaluation metrics used for our experimental setup.

\subsection{Dataset} \label{ssec:dataset}
We demonstrate the performance of our VAD method on the \textbf{Aurora-2} dataset \cite{aurora}, which contains clean and synthetically degraded speech utterances. Clean speech utterances are borrowed from the TIDigits dataset \cite{leonard1984database}, downsampled at $8$ kHz with ideal low--pass filter characteristics, to which noise is added from the NOISEX-92 dataset \cite{journals/speech/VargaS93}. The overall data is distributed in $4$ directories as train (\textit{Train}), and test sets (\textit{Test A, Test B, and Test C}). The \textit{Train} set contains $8440$ utterances, with a mix of clean signals, and signals degraded with `train', `car', `babble', and `exhibition hall' noises. The \textit{Test A} set contains different utterances subjected to a channel characteristics and degraded with same noise types, at SNR levels of 20, 15, 10, 5, 0, and -5 dB. The \textit{Test B} and \textit{Test C} sets contain signals degraded with different noise types and channel characteristics, to present a disjoint environment from training. Together the three test sets contain 4004 utterances. In total, the train and test sets contain 8440 and 70070 audio files respectively. The labels are obtained from \cite{TAN20201}, generated using a HTK recognizer \cite{Young2006}, trained on $12$ MFCC coefficients, $\Delta$+$\Delta\Delta$s, and log-energy, computed over the \textit{Train} set, modeled by $16$ HMM states, each represented by 3 Gaussian mixtures.

\subsection{Baseline Methods} \label{ssec:baselines}
Performance of the proposed method ($V_{\text{ZFF}}$) is compared against several supervised and unsupervised methods in the literature, implemented based on the hyper-parameters values given in their respective papers. \\
\textbf{rVAD} ($V_{\text{RVP}}$) \cite{TAN20201}: implements two passes of denoising as enhancement to high energy speech regions: \textit{i)} an \textit{a-posteriori} weighted energy difference measure, subjected to a pitch detection routine, used to classify speech from noise; \textit{ii)} a spectral subtraction method for speech enhancement. The VAD stage uses a SNR weighted energy measure along with the pitch information to determine voice segments. \\
\textbf{rVAD--Fast} ($V_{\text{RVS}}$) \cite{TAN20201}: a faster implementation which uses spectral flatness as a measure to identify the presence of pitch in a segment. \\
\textbf{VAD--Wavlet} ($V_{\text{DWT1}}$ and $V_{\text{DWT2}}$) \cite{dwt-code}: uses detail coefficient in wavelet based decomposition of speech. Daubechie's wavelet ($dB3$) is used to derive details at multiple levels. Two different methods are implemented: \textit{i)} $V_{\text{DWT1}}$, which uses RMS energy of details coefficients to discriminate between speech and noise, and \textit{ii)} $V_{\text{DWT2}}$, which uses four energy based parameters. \\
\textbf{VAD--Fusion}($V_{\text{FUS}}$) \cite{van2013robust}: a MLP based voice/non-voice classifier implemented over a fusion of multiple spectral features derived by exploiting the spectro-temporal modulations, harmonicity, and long term spectral variability of signal. \\
\textbf{VAD--LTSD} ($V_{\text{LTSD}}$) \cite{ramirez2004efficient}: compares long term spectral envelope characteristics of a segment against average noise spectrum. An adaptive threshold updates over noise power which is estimated after each non-speech segment is discovered. \\
\textbf{GP--VAD} ($V_{\text{GP}}$) \cite{Dinkel2020}: a convolutional recurrent neural network (CRNN) trained on noisy log-Mel power spectrograms in a weakly supervised fashion using only clip-level labels. \\
\textbf{VAD--TEO}  ($V_{\text{TEO}}$) \cite{hegde2019voice}: highlights formant information using a band-spectral mass function, derived from a convex spectral energy function, used to compute the spectral entropy to determine voiced/unvoiced regions. \\
\textbf{VAD--LSD} ($V_{\text{LSD}}$) \cite{ramirez2004efficient}: uses maximal spectrum information to model the contrast within spectral behavior across voicing and noise segments, and derives average noise spectral characteristic from silence and pause regions to obtain a divergence function. \\
\textbf{VAD--LSE} ($V_{\text{LSE}}$) \cite{pang2017spectrum}: compares the energy content within high and low spectral bands in the DFT spectra, knowing that voicing information is predominantly confined in the latter band.
\begin{figure*}[ht]
  \centering
  \includegraphics[width=\linewidth]{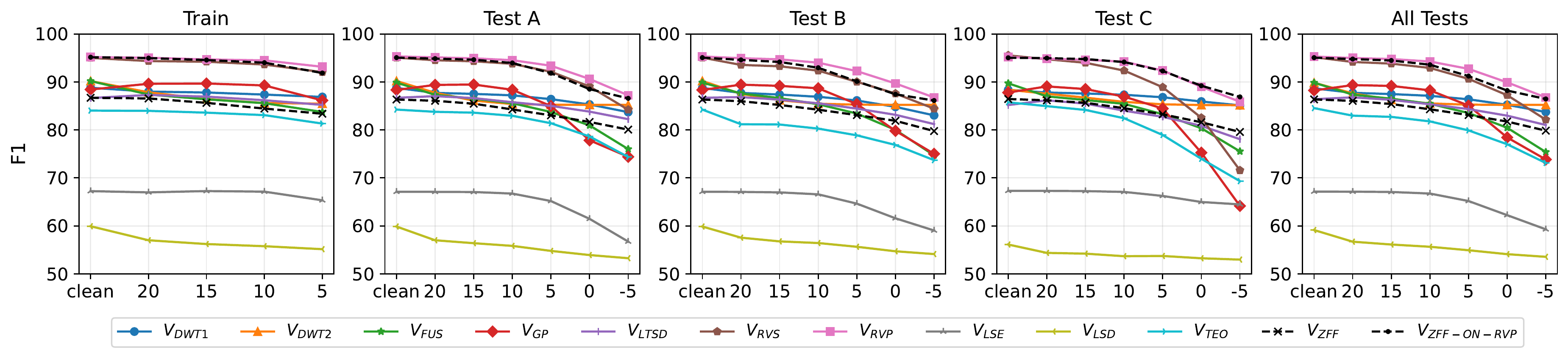}
  \caption{Performance of methods across all SNRs in different sets of the Aurora-2 database.}
  \label{fig:f1_scores}
\end{figure*}
\subsection{Evaluation Metrics} \label{ssec:metrics}
VAD can be treated as a binary classification problem of sorting the input signal frames into voiced/non-voiced classes. To that end, frame-level results such as true positives (TP), false positives (FP), false negative (FN), and false positives (FP) can be used to compute standard classification metrics and thus measure a model's performance over time. We use precision (P), recall (R), and F1-score, which are calculated as given in \cref{eqn:metrics}.
\begin{equation}
   \text{P} = \frac{\text{TP}}{\text{TP}+\text{FP}}; \quad
   \text{R} = \frac{\text{TP}}{\text{TP}+\text{FN}}; \quad
   \text{F1} = 2 \cdot \frac{\text{P} \cdot \text{R}}{\text{P} + \text{R}}
   \label{eqn:metrics}
\end{equation}

\section{Results and Discussion} \label{sec:results}

\Cref{fig:f1_scores} shows the performance of the different VAD methods. V$_{\text{ZFF}}$ refers to the method based on the first approach, presented in~\Cref{sec:method}. V$_{\text{ZFF-on-RVP}}$ denotes the second approach, where the composite signal $r_c$ is fed as input to rVAD algorithm \emph{without application of the denoising routine}. The results show that V$_{\text{ZFF}}$, which uses minimal spectral information, outperforms $V_{\text{TEO}}$, $V_{\text{LSD}}$ and $V_{\text{LSE}}$, and achieves a performance close to most of the other supervised and unsupervised methods, with the exception of $V_{\text{RVP}}$. It is also interesting to observe that V$_{\text{ZFF-on-RVP}}$, computed without the denoising routine, yields performance close to $V_{\text{RVP}}$. Furthermore, it can also observed that at very high and very low SNRs, the V$_{\text{ZFF}}$ and V$_{\text{ZFF-on-RVP}}$ methods yield competing performances. Together these observations show that the outputs of zero-frequency filters indeed carry the source and system information in a reliable and robust manner, and can be effectively employed for VAD.

\Cref{table:stddevs} shows the standard deviation of the F1-scores of each method, across all SNRs for the entire \textit{Test} set. It can be observed that performance of V$_{\text{ZFF}}$ remains invariant to added interferences across a range of SNRs ($20$ dB to $-5$ dB).  
\Cref{fig:f1_scores} shows that the performance of the method not only suffers very marginally at low SNRs, but in fact gives good results for the very low SNR value of $-5$ dB. This is particularly noticeable for the \textit{Test C} set, where the channel characteristics are different than in the \textit{Train} set (G.712). Nonetheless, $V_{\text{ZFF}}$ outperforms $V_{\text{GP}}$, $V_{\text{RVS}}$, $V_{\text{FUS}}$, and $V_{\text{LTSD}}$.

\begin{table}[!htb]
    \caption{The standard deviation of the F1-scores ([\%]) for each method, computed over the entire \textit{test} set and across all SNRs. A high value indicates a significant variance and degradation with noise.}
    \begin{adjustbox}{width=\columnwidth}
    \label{table:stddevs}
    \centering
    \begin{tabular}{ccccccccccc}
    \toprule
    $V_{\text{DWT}}$ & $V_{\text{LSD}}$ & $V_{\text{LTSD}}$ & $V_{\text{ZFF}}$ & $V_{\text{LSE}}$ & $V_{\text{RVP}}$ & $V_{\text{ZFF-ON-RVP}}$ & $V_{\text{TEO}}$ & $V_{\text{RVS}}$ & $V_{\text{FUS}}$ & $V_{\text{GP}}$ \\ 
    1.6 & 1.7 & 2.0 & 2.2 & 2.8 & 3.0 & 3.2 & 3.7 & 4.3 & 4.5 & 5.7 \\
    \bottomrule
    \end{tabular}
    \end{adjustbox}
\end{table}

\begin{figure}[htb]
\begin{minipage}{\linewidth}
  \centering
  \includegraphics[width=\textwidth]{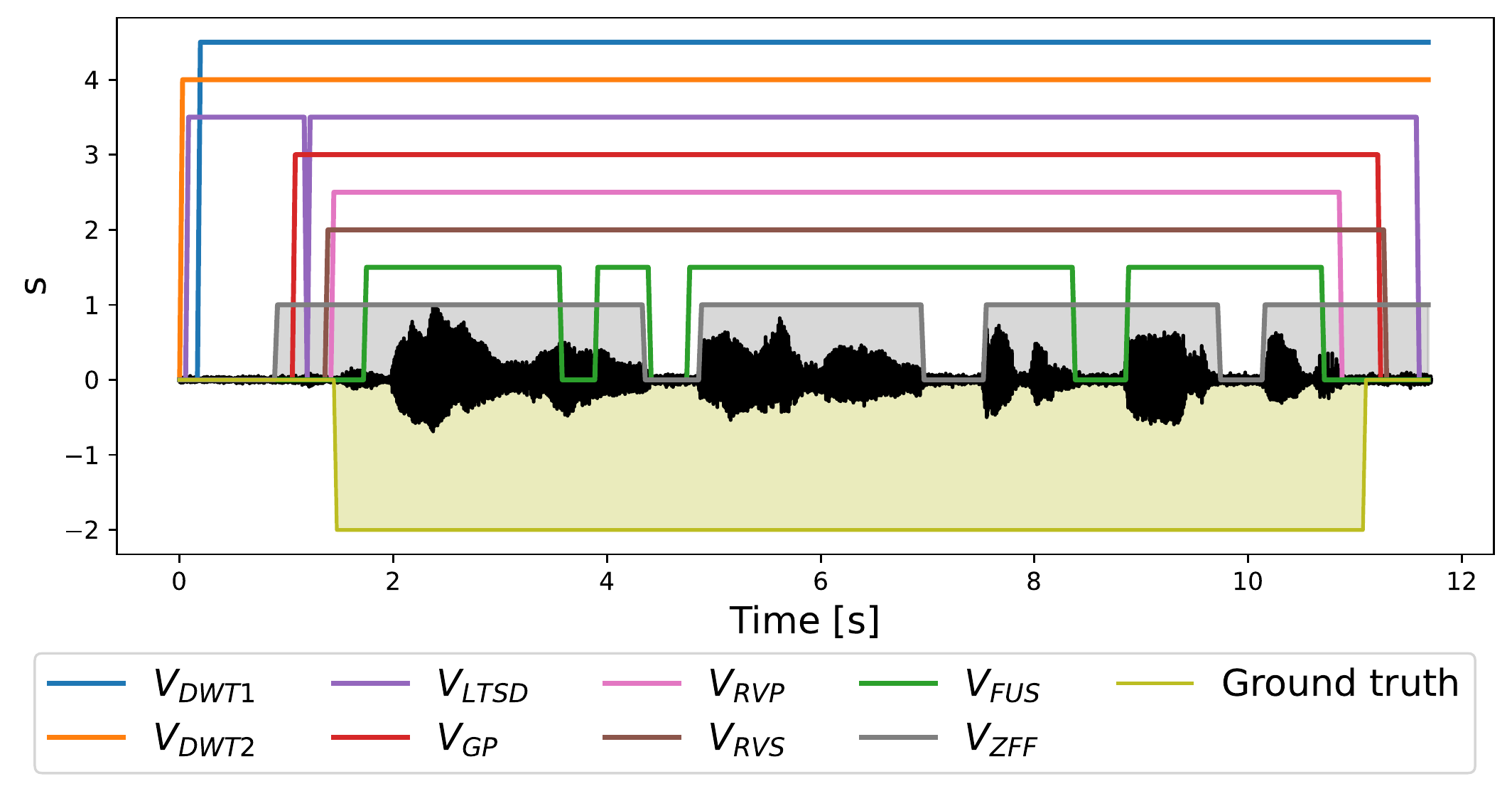}
  \caption{Performance of some baseline methods, proposed method, and ground truth, for noisy speech (SNR = 10 dB).}
  \label{fig:scores_all_methods}
\end{minipage}
\end{figure}

\Cref{fig:scores_all_methods} shows the decision boundaries obtained for different methods on a given speech recording. It can be observed that the decision boundary produced by $V_{\text{ZFF}}$ is successfully able to segment the speech signal into intervals which are significantly tighter than the other methods, as well as those given in the ground truth. In other words, the performance of our method is not as well reflected in the F1-scores compared to the other methods because of the broader boundaries of the ground truth segments, even though our method is able to provide a much more granular segmentation. The methods yielding higher performance comply closely to the broader VAD boundaries and consequently yield higher F1-scores, as can be noted in \Cref{fig:scores_all_methods} for $V_{\text{RVP}}$, $V_{\text{RVS}}$, $V_{\text{DWT1}}$, $V_{\text{DWT2}}$, and $V_{\text{GP}}$ methods.  


\section{Conclusion} \label{sec:conclusion}
In this paper, we investigated modeling source and system information jointly using zero-frequency filtering technique for voice activity detection. In that direction, we proposed and validated two approaches for VAD on the Aurora-2 dataset with different noise, channel, and SNR conditions. Our investigations demonstrated that VAD can be effectively performed by combining the filter outputs together to compose a composite signal carrying $f_0$, $F_1$, and $F_2$ related information, and then applying a dynamic threshold after spectral entropy-based weighting (first approach), or else by passing the composite signal to another VAD (second approach). The experiments also illustrate that the proposed method produces more refined boundary demarcations for the VAD task compared to other supervised and unsupervised methods in the literature. It is also robust against degradation as well as channel characteristics, and yields stable performance across a range of SNRs. The first approach operates in the time domain and is relatively less complex to implement. The second approach illustrates that the composite signal, obtained by modulation of trend removal in the zero-frequency filtering, is an effective representation of speech characteristics, and can hence be used in conjunction with other VADs.

One of the main advantages of the proposed zero-frequency filtering based approach is that it does not explicitly assume any mathematical model for the produced speech signal in order to acquire source and system information. It can also thus be extended to other types of audio signals, such as animal and birds vocalizations. Our future work will focus in this direction, along with modeling the composite signal using the raw waveform neural network based modeling approach~\cite{Palaz_INTERSPEECH_2013} for supervised voice activity detection~\cite{ZazoIS16}.


\section{Acknowledgement}
This work was partially funded by the Swiss National Science Foundation (SNSF) through projects: NCCR Evolving language (grant agreement no. 51NF40\_180888) and Towards Integrated processing of Physiological and Speech signals (TIPS) (grant agreement no. 200021\_188754).

\bibliographystyle{IEEEtran}
\bibliography{mybib}

\end{document}